# Exploring Runtime Evolution in Android: A Cross-Version Analysis and Its Implications for Memory Forensics

Babangida Bappah, Lauren G Bristol, Lamine Noureddine, Sideeq Bello, Umar Farooq, and Aisha Ali-Gombe
Louisiana State University, USA
bbappa1@lsu.edu, lbrist8@lsu.edu, lnoureddine@lsu.edu, sbell49@lsu.edu, ufarooq@lsu.edu, aaligombe@lsu.edu

*Abstract*—Userland memory forensics has become a critical component of smartphone investigations and incident response, enabling the recovery of volatile evidence such as deleted messages from end-to-end encrypted apps and cryptocurrency transactions. However, these forensics tools, particularly on Android, face significant challenges in adapting to different versions and maintaining reliability over time due to the constant evolution of low-level structures critical for evidence recovery and reconstruction. Structural changes, ranging from simple offset modifications to complete architectural redesigns, pose substantial maintenance and adaptability issues for forensic tools that rely on precise structure interpretation. Thus, this paper presents the first systematic study of Android Runtime (ART) structural evolution and its implications for memory forensics. We conduct an empirical analysis of critical Android runtime structures, examining their evolution across six versions for four different architectures. Our findings reveal that over 73.2% of structure members underwent positional changes, significantly affecting the adaptability and reliability of memory forensic tools. Further analysis of core components such as Runtime, Thread, and Heap structures highlights distinct evolution patterns and their impact on critical forensic operations, including thread state enumeration, memory mapping, and object reconstruction. These results demonstrate that traditional approaches relying on static structure definitions and symbol-based methods, while historically reliable, are increasingly unsustainable on their own. We recommend that memory forensic tools in general and Android in particular evolve toward hybrid approaches that retain the validation strength of symbolic methods while integrating automated structure inference, version-aware parsing, and redundant analysis strategies. These adaptations are essential for sustaining effective and trustworthy forensic capabilities amidst rapidly evolving runtime environments.

## I. INTRODUCTION

Memory forensics has become increasingly vital in digital investigations, especially now that smartphones are the primary source of digital evidence. While traditional disk-based forensics remains important, memory forensics has emerged as an essential capability for understanding system and application runtime behavior, enabling the recovery of volatile evidence that may not persist on disk. For memory forensics tools to be effective, they must accurately parse memory images and extract meaningful information. This requires precise identification, recovery, and interpretation of in-memory data structures, which depend on an exact understanding of their layouts and member locations. Even a single misaligned offset can invalidate the entire analysis, leading to inaccurate or incomplete results.

This assumption breaks down rapidly on Android. Android's architecture adds a complex Android runtime (ART) on top of the Linux kernel, managing memory allocation, thread lifecycles, object instantiation, and execution behavior for user-facing apps. This runtime evolves with each new Android release: structure layouts change, internal optimizations are introduced, and previously accessible components are reorganized or removed. These changes are often undocumented, deeply implementation-specific, and tightly coupled to both Android version and hardware architecture. As a result, forensic tools that rely on static assumptions about structure layouts – such as symbol-based or profile-based analyzers – routinely fail to operate on newer versions of the platform without significant manual intervention.

This volatility in Android's internal memory structures and runtime behavior, driven by frequent changes across Android OS versions presents challenges that differ sharply from those encountered in traditional Linux memory forensics. In Linux, kernel data structures are often better documented and benefit from long-standing conventions around debug symbol generation and versioning. Android, by contrast, introduces a faster, more fragmented, and less transparent evolution cycle, further complicated by vendor-specific builds, support for multiple ABIs (e.g., ARM32/64, x86/64), and varying build configurations (e.g., debug vs. release). These factors make it difficult to generalize or reuse structure layouts across devices or versions, leading to brittle forensic pipelines and increasing the time and expertise required for each analysis.

Symbol-based approaches, also known as profile-based approaches, form the foundation of most memory forensics tools. These approaches rely on debugging symbols to extract structure layouts and create profiles containing memory locations of crucial data structures. Kernel-level memory forensics tools like Volatility[1] and Rekall[2] and Userland memory forensics tools like DroidScraper[3] use these profiles to recover forensic artifacts of interest, producing consistent, defensible results that can withstand scrutiny in investigative and legal contexts. However, as technology evolves rapidly, maintaining accurate profiles for each operating system version becomes increasingly impractical, requiring continuous updates that often lag behind new releases. This challenge of structural evolution is particularly prominent in Android due to its rapid release cycles and the widespread deployment of different versions. With each new version, structure members may be repositioned, core structures removed, and new ones introduced. These changes force forensic tools to either continuously adapt to the evolving data structures in new versions or fail to keep up the maintenance pace, eventually becoming obsolete.

Recent research has proposed symbol-agnostic methods which attempt to infer structure layouts through heuristics, pointer analysis, or dynamic inspection, as an alternative to symbol-based methods. For instance, OS-agnostic methods based on pointer topology [4] and code-based structure inference [5] aim to bypass static assumptions by dynamically recovering layout information. While these approaches show promise for automation, they often lack semantic guarantees and are difficult to validate – limitations that are particularly problematic in forensic contexts where reproducibility and evidentiary rigor are paramount. As a result, symbol-based methods remain the standard in Android memory forensics, valued for their correctness, traceability, and compatibility with legally defensible workflows, despite their poor scalability across evolving runtime versions.

Yet, this reliance on symbol-based techniques raises a critical question: *How sustainable is the continued dependence on symbol extraction, given the effort required to identify and adapt structure definitions across different OS versions?* This challenge is especially pronounced in operating systems like Android, which feature a layered architecture that places a complex runtime environment on top of the Linux kernel, further complicating analysis. To extract reliable forensic evidence from Android, forensic tools must accurately interpret not only low-level OS structures but also complex, unique, and often newly introduced userland data structures.

Thus, in this paper, to the best of our knowledge, we present the first comprehensive measurement study examining the practical challenges of extending Android memory forensic tools from older to newer Android versions. By conducting cross-version binary profiling and symbol-based structural analysis of Android Runtime across six versions (9–14) and four architectures (ARM32, ARM64, x86, and x64), we quantify both the extent and nature of structural evolution. Our findings reveal significant changes in key memory forensics data structures such as Thread, Heap, and Runtime, including size variations, member additions, and removals. Newer versions introduce features like interpreter cache and generational garbage collection that contribute to increased complexity. These structural modifications have critical implications for forensic tool development, maintenance, and reliability, underscoring the need for adaptable methodologies to address the challenges posed by Android's rapid evolution.

To summarize, our research makes salient contributions:

- We present the first systematic and comprehensive analysis of Android Runtime structure evolution, examining changes across six major versions for four architectures.
- We quantify the impact of structural changes on key forensic tasks, demonstrating how these modifications influence critical capabilities such as thread enumeration, heap analysis, and object recovery.
- We provide a publicly accessible repository containing detailed runtime structure information for Android versions 9 to 14, encompassing ARM and x86 for 32—and 64-bit architectures. This resource will assist forensic tool developers and researchers in advancing Android memory analysis research [6]
- We propose recommendations for developing adaptable memory forensic tools that can effectively address the challenges posed by Android Runtime evolution.

This paper is structured as follows: Section II provides background and related work on the Android Runtime, as well as memory forensics techniques and challenges. Section III details our methodology for collecting and analyzing structural changes across Android versions. Section IV presents our evaluation and findings on runtime evolution and forensic implications. Section V discusses its impact on forensic tool reliability. Section VI provides recommendations for forensic tool development. Finally, Section VII concludes with key contributions and future directions.

## II. BACKGROUND, RELATED WORK AND CHALLENGES

In this section, we provide a concise background of Android's runtime architecture and its implications for Android memory forensics. We then review how existing forensic approaches have attempted to handle the rapid evolution of Android's internal memory structures, highlighting their limitations. Finally we outline the practical challenges of extending forensic tools across the rapid evolution of Android versions.

### A. Android Memory Forensics

In mobile device architectures such as Android, where applications operate in isolated runtime environments with copies of shared libraries and dedicated memory regions, **Userland memory forensics** is more suited for extracting user-level and application-specific evidence. This approach allows for the recovery of volatile data, such as user interactions and app states, that may not be accessible through kernel-level analysis. In recent years, the community has made significant strides in designing userland memory forensics tools targeting the Android runtime. The runtime environment serves as a software layer that provides essential services, tools, and resources for executing programs or applications. Specifically on Android, the runtime acts as an intermediary between applications and the underlying hardware or operating system, managing tasks such as memory allocation, I/O, and process scheduling to ensure seamless execution. Since its inception, the Android runtime has undergone substantial evolution, introducing new complexities and challenges for forensic analysis. The transition from the Dalvik Virtual Machine to the Android Runtime (ART) in Android 5.0 fundamentally changed how applications are executed, shifting from a register-based interpretation model with just-in-time (JIT) compilation to ahead-of-time (AOT) compilation with advanced optimization capabilities. Early works such as [7] and [8] introduced the first userland memory forensics approaches for examining the contents of the Android Dalvik Virtual Machine. Building on this foundation, GUITAR [9] advanced GUI reconstruction by implementing methods to recover deallocated interface elements, while VCR [10] introduced app-agnostic evidence recovery by targeting Android's intermediate service architecture.

After transitioning to the new ART runtime, Android introduced the Runs-Of-Slots Allocator (RosAlloc) memory allocation algorithm, which, along with the unique characteristics of the new runtime, impacted the adaptability of existing forensic tools. This allocator organizes heap memory into rows of slots of uniform size, grouping them into pages within brackets. The first page of each bracket contains a header that specifies the number of pages in the bracket and maintains a bitmap for slot allocation. The number of slots per page is determined based on the bracket's size, header length, and device architecture-specific byte alignment requirements. Each slot stores the data for a single object, with the initial bytes containing the address of the object's parent class. Objects are classified by size to minimize fragmentation and facilitate parallel garbage collection, while larger objects exceeding 12 KiB are allocated in Large Object Space (LOS) areas. RosAlloc uses a page map in RAM to manage bracket pages (4 KiB each), with heap space typically starting near the process's lowest virtual address, commonly at 0x12c00000. In 2017, Soares [11] introduced the first memory forensics technique for extracting and analyzing data allocated using the RosAlloc memory management scheme in the Android ART runtime. This recovery technique was further extended into Timeliner [12] in 2018, where the authors demonstrated temporal reconstruction capabilities. By analyzing residual data structures in Android's ActivityManagerService, Timeliner enabled the recovery of user action sequences from memory images.

Between Android 5 and 7, the evolution of garbage collection in ART progressed through four algorithms: Semi-Space, Generational Semi-Space, Concurrent Mark Sweep (CMS), and Concurrent Copying (CC)[13, 14]. With the release of Android 8, Google introduced a significant change to ART's garbage collection by adopting the CC as the default collector and transitioning to the Region Space Allocation memory allocation algorithm, rendering many existing forensic techniques obsolete. In region space allocation, objects are allocated to specific regions and collected based on live object thresholds. The runtime further optimizes garbage collection by employing distinct strategies for foreground and background processes, with dedicated collectors tailored to each scenario. To address the challenges posed by these changes in runtime technology, [3, 15, 16] introduced userland memory forensic techniques specifically designed for the recovery and reconstruction of objects allocated using the region-based memory management approach.

*B. Related Work*

Memory forensics has evolved significantly over the past decade, with Case and Richard[17] chronicling its progression from basic memory analysis to sophisticated runtime investigation techniques. By design, it operates at two distinct levels: kernel space (traditional) and userland. While kernel-level memory forensics focuses on recovering data and evidence from low-level system structures and operations, userland memory forensics examines application-specific memory regions, providing direct access to application runtime behavior and user activities.

In the last two decades, traditional or kernel-level memory forensics has seen significant advancement in tools and techniques. The FATKit framework[18] pioneered systematic structure extraction by automatically deriving object definitions from source code. Following this, several works addressed the challenge of structure information acquisition. The Volatility Framework, established in 2007, is an open-source memory forensics platform widely used by professionals and researchers to investigate cybercrime, malware, and system intrusions by analyzing memory images from various operating systems[1, 19, 20]. Through symbol identification and kernel-level data structure definitions, Volatility supports tasks such as process enumeration, network connection analysis, and the recovery of hidden or malicious activity in memory. Numerous research efforts in this domain have proposed plugins to extend Volatility's capabilities[21, 22]. Moving beyond kernel-level analysis, Block and Dewald[23] made significant contributions by focusing on user-space heap memory in Linux processes, particularly examining Glibc's ptmalloc2 implementation to extract valuable forensic artifacts like credentials and application state. The same authors later developed techniques for detecting hidden injected code in Windows memory by analyzing Page Table Entries rather than relying on VAD structures[24]. As artificial intelligence becomes increasingly prevalent, Oygenblik et al.[25] introduced AiP, a framework for recovering and rehosting deep learning models from memory images, enabling white-box inspection for security validation. As an alternative to static data structure identification, Lin et al.'s SigGraph [26] introduced graph-based signatures for kernel structure identification, while Feng et al.'s ORIGEN [27] developed methods to extract offset-revealing instructions for cross-version analysis. Cohen's work [28] characterized the variability in Windows kernel versions and its impact on memory analysis accuracy. To address the lack of debug information, alternative approaches emerged, with LogicMEM [29] demonstrating profile generation through logic inference and AutoProfile [30] automating profile generation using dynamic analysis. Taking a different direction, Oliveri et al. [4] proposed FOSSIL, an OS-agnostic approach identifying data structures based solely on topological properties rather than OS-specific models, enabling forensic analysis across different operating systems without prior knowledge of kernel internals. Franzen et al.[5] introduced Katana, which reconstruct Linux structure layouts through code-based analysis using Ghidra's architecture-independent P-code, effectively handling configuration variation and structure layout randomization across different architectures. Saltaformaggio et al.[31] developed DSCRETE, which identifies and reuses applications' own rendering logic to interpret application-specific encoded data structures in memory dumps, providing human-understandable output for forensic analysis. Song et al.[32] created DeepMem, applying deep learning techniques to memory forensics by converting memory dumps into memory graphs and using graph neural networks to detect kernel objects without OS-specific knowledge. Zhang et al.[33] presented RAMAnalyzer, which automatically analyzes Linux memory images without prior kernel version information through a five-stage pipeline that dynamically reconstructs kernel code to extract live system information. While these works offer alternative approaches to symbol-based memory forensic analysis, they primarily focus on kernel-level structures. Moreover, although they may present more practical solutions, these alternatives do not provide the level of reliability and validation required for forensic analysis methodologies. As a result, they have yet to be adopted by any major analysis tool or engine.

### C. Memory Forensics Challenges

Changes in Android's runtime architecture and the limitations of existing approaches to ensure reliability and validation, underscore the growing challenges faced by forensic tools. Although Android has retained ART as the runtime and adopted CC and Region Space Allocation for garbage collection and memory management since Android 8, the design and implementation of runtime management particularly the data structures continue to evolve. As these structures change, they impact the reliability and adaptability of existing forensic tools, even when the underlying runtime and memory allocation mechanisms remain unchanged. For instance, the Android ART manages active `Threads` using a double link list. The pointer to this link list exists as a field at a specific offset within the `ThreadList` structure, which has its pointer in the `Runtime` structure. Additionally, the thread name, ID, and metadata must be properly dereferenced from their offsets to recover and reconstruct the correct information. However, any changes to the structures, data types, or offset locations of the structure or its members in a new version can significantly affect the adaptability and maintenance of memory forensic tools. Another source of variability in Android runtime structures is the heterogeneity of Android hardware, which is determined by the compilation options used when building the Android image. The Android AOSP currently supports more than 10 compilation options for each version. Device images can be compiled for various architectures, including ARM, x86, MIPS, and others, with support for both 32-bit and 64-bit versions. In addition to these fundamental architectural differences, Android allows for builds in multiple modes, such as debug, user, eng, and release. These build modes can significantly affect the runtime data structures' number, types, and sizes. Lastly, manufacturers' customizations add another layer of complexity. OEMs like Samsung often modify or tune these data structures to support proprietary functionalities and features. These customizations further increase variability, making it challenging for tools to rely on fixed or precise assumptions about runtime structure layouts. Given these enumerated challenges, one can hypothesize that they pose significant issues for memory forensics. However, to date, no empirical measurements have been conducted to assess the depth and impact of these modifications. As such, this work seeks to evaluate the macro- and micro-level trends in the current Android Runtime and their implications for userland memory forensics. To this end, we pose the following Research Questions (RQs):

**RQ1** How do Android version updates and architecture differences influence the size of the ART library?
**RQ2** What trends can be observed in the size of debugging symbols across the different architectures and Android versions?
**RQ3** What broad structural changes occurred across Android versions for critical memory forensics data structures?
**RQ4** How do core runtime structures, such as Runtime, Thread, and Heap, evolve across Android versions?
**RQ5** What is the impact of structural changes on critical memory forensic tasks?

### III. STUDY DESIGN, DATA COLLECTION AND ANALYSIS

To answer the research questions posed in Section II-C and systematically study the variability in Android versions and the evolution of their structures, particularly those related to memory forensics, we first develop a process for data collection and then adopt two analysis methods for the evaluation - Cross-Version Binary Profiling and Symbol-Based Structural Evolution Analysis.

### A. Data Collection

For this study, we analyzed Android versions 9 through 14 to capture both the historical context and the significance of newer technological advancements. Android 15, the latest stable version, was not included as it had not been released when this study commenced. For all analyzed versions, both 32-bit and 64-bit variants were examined for ARM and x86 architectures. The data collection process extracts key binary characteristics, debugging symbol statistics, and other relevant memory forensic data structures, ensuring consistency and reproducibility across all versions.

*1) Preparing the Build Environment:* The AOSP build process requires specific environment setup and dependencies. A Linux-based system with minimum requirements of 250GB disk space and 8GB RAM is necessary for building the source code[34]. Our builds were conducted on a 13th Gen Intel i9-13900KF processor with 64GB RAM and 4TB storage, with each version taking approximately 4-5 hours to compile. The built environment requires the installation of essential tools, including Git, Python, OpenJDK, and the Repo tool for source synchronization. For each version, we set up separate, clean build environments for documentation to ensure the reliability and reproducibility of the process.

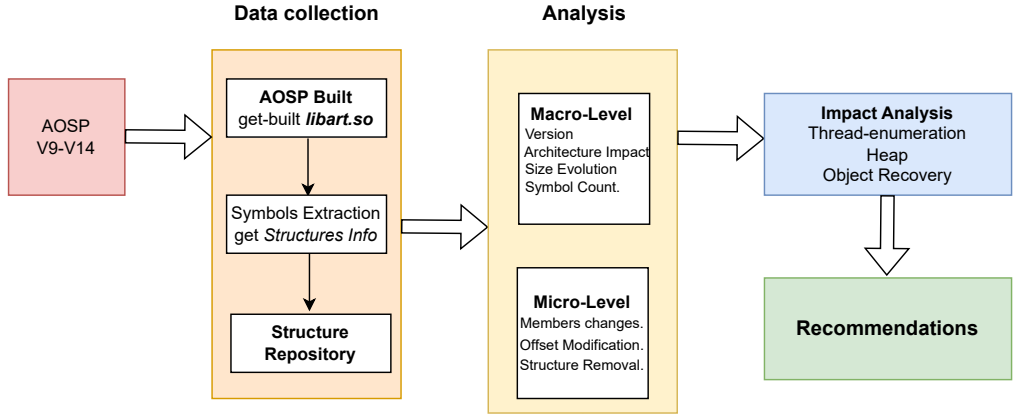

Fig. 1: Study Framework for Analyzing the Evolution of Android Runtime (ART) Structures and their Impact on Memory Forensic Methodologies

*2) Build Configuration and Compilation:* The build configurations were tailored to the target architectures (ARM and x86) using the *lunch* command, with both 32-bit and 64-bit variants analyzed for all versions (9-14). We select the 'eng' build variant, which is designed for faster builds and day-to-day development. This variant provides additional flexibility for system-level development and enables debug capabilities essential for our analysis. After the build config, we compiled the source code for each version to produce system images. The compilation process automatically resolved inter-module dependencies optimized the binaries for the selected architectures, and produced the final shared libraries and executables. These output images included the key ART library (libart.so), which serves as the primary focus of our analysis.

*3) Extracting DWARF Symbols:* Following the AOSP build process, we extracted detailed structure information from the compiled libart.so for each version and architecture combination. Extracting this information presents unique challenges due to the Android Runtime's C++ implementation, which extensively uses inheritance hierarchies, templates, and complex object-oriented patterns, thus making it difficult to use existing DWARF carving tools such as the Dwarf2Json[35]. To address these challenges, we developed a custom parser tailored to handle the specific complexities of the Android Runtime. Using the pyelftools library[36], our parser processes DWARF information from the DWARFInfo context object that is derived from the ELF file to extract precise structure layouts. It begins by iterating through each Compilation Unit (CU), identifying structure definitions using DWARF tags. Starting at the CU, the parser locates and iterates through its Debugging Information Entries (DIEs). For each DIE, objects are identified using the tags DW_TAG_class_type, DW_TAG_structure_type, and DW_TAG_member. For entries with these tags, the parser retrieves the name (DW_AT_name) and the size in bytes (DW_AT_size). The parser then examines the children of each object, if present. If a child entry has a tag identifying it as an object and contains a non-null location attribute (DW_AT_data_member_location), it is determined not to be a constant, which aligns with our focus on objects and class members. For such entries, the parser retrieves the child's name and location within the parent object (DW_AT_data_member_location). Entries without a name are defaulted to "UnNamed." Thus, traversing through the entire debug section of each libart.so, for each structure found, the parser will record its total size and analyze its members, capturing the name and precise offset of each member within the parent structure. This approach ensures an accurate representation of runtime data structures, including their memory layouts, as they exist both at compilation and runtime. Using the custom parser, we extracted all DWARF debug symbols embedded within each compiled libart.so. The extracted information, including structure names, total sizes, and member offsets, is stored in a structured JSON format for each version and architecture. This standardized format facilitates cross-version comparison and analysis. To ensure comprehensive documentation, all version-specific structure information is maintained in a centralized repository for Android versions 9 to 14 across ARM and x86 architectures.

*B. Analytical Methods*

The data collected, including binary characteristics, profiling data, and DWARF symbols for all analyzed versions and architectures, is passed to the analysis phase, as shown in Figure 1. For our measurement, We employ two analytical methods: Cross-Version Binary Profiling and Symbol-Based Structural Evolution Analysis.

*1) Cross-Version Binary Profiling:* The cross-version binary profiling examines key metrics across different Android ART library versions and architecture to identify macro-trends in system evolution. Our primary objective is to leverage the binary characteristics collected in our analysis to assess the impact of Android versioning and architecture on overall runtime growth and complexity. For this analysis, we focus on Research Questions 1–2 outlined in Section II.

*2) Symbol-Based Structural Evolution Analysis:* While the cross-version binary profiling focuses on macro-level trends, the symbol-based structural evolution analysis focuses on micro-level changes. This analysis examines the evolution of the runtime data structures by parsing the specifics of each debug symbols extracted from the ART library. We focus on identifying and categorizing structural differences, such as size variations, member additions, and removals, to understand the impact of these changes on memory layout and runtime behavior. Specifically, we aim to address research questions 3-5 posed in Section II.

To answer these research questions, we developed another custom differing tool to compare the extracted symbols across versions. For each version pair, we analyze JSON files containing detailed structure information, systematically identifying changes in the structure composition and layout. Our comparison identifies three primary types of structure changes: (1) added structures (present in the current version but not in the previous), (2) removed structures (present in a previous version but not in the current), and (3) modified structures (present in both versions but with changes). We perform a detailed analysis of member-level changes for modified structures, categorizing them into additions, removals, and offset modifications. For each version transition, we store the comparison results in a structured JSON format documenting: Added and removed structures; Member-level changes (additions, removals, offset modifications); Structure size changes; Affected member offsets with both old and new locations.

Due to the extensive number of structures in the ART library (over 19,000 DWARF symbols in recent versions), we focused our analysis on 34 critical data structures essential for memory forensics. We identified these structures by examining DroidScraper[3], a stable Android memory forensic tool supporting Android versions 8, 8.1 and 9, and determining which low-level structures would be necessary to extend its functionality to newer Android versions. These critical components include core structures like Thread, Runtime, and Heap, along with supporting structures required for thread enumeration, memory mapping, and object recovery.

## IV. ANALYSIS AND EVALUATION

In this section, we present the results of the two data analysis methods described in Section III to evaluate the evolution of Android Runtime structures across versions and architecture, and their impact on memory forensic capabilities.

### A. Cross-Version Binary Profiling

Table I presents the results of our analysis of the ART library across Android versions 9 through 14 for four different architectures (ARM32, ARM64, x86 and x64), highlighting changes in binary size and DWARF debugging symbols. This data provides insights into runtime evolution patterns and architectural impacts, which inform our subsequent structural analysis.

| (a) ART Library Size (MB) | | | | |
|---|---|---|---|---|
| Version | ARM64 | x86_64 | ARM32 | x86_32 |
| 9  | 157.57 | 156.05 | 184.41 | 189.80 |
| 10 | 147.42 | 147.36 | 128.77 | 133.43 |
| 11 | 198.32 | 200.71 | 162.51 | 167.94 |
| 12 | 177.09 | 172.35 | 153.99 | 157.29 |
| 13 | 89.76  | 88.70  | 83.90  | 88.02  |
| 14 | 92.52  | 92.08  | 86.53  | 91.36  |

| (b) Debugging Symbols Count | | | | |
|---|---|---|---|---|
| Version | ARM64 | x86_64 | ARM32 | x86_32 |
| 9  | 15,698 | 15,698 | 15,702 | 15,731 |
| 10 | 16,757 | 16,758 | 16,789 | 16,800 |
| 11 | 17,215 | 17,202 | 17,238 | 17,227 |
| 12 | 17,743 | 17,746 | 17,744 | 17,775 |
| 13 | 17,275 | 17,270 | 17,228 | 17,262 |
| 14 | 19,115 | 19,100 | 19,107 | 19,097 |

TABLE I: Evolution of ART Library Across Android Versions

*RQ1: Changes in Binary Size:* Our analysis of the ART binary characteristics shows significant fluctuations in size across versions. The binary size shows notable variation, peaking in Android 11 at approximately 200MB for x86_64 architecture - a 36% increase from Android 10, as shown in Table I(a). This substantial growth suggests a significant expansion in runtime capabilities and features during this period. However, we observe a dramatic shift in Android 13, where the library size shrinks by approximately 48% compared to Android 12 across all architectures (e.g., from 177MB for 12 to 89MB in 13 on ARM64). Using **readelf**, we examined the binaries and found that the debug sections for Android 13 and 14 are encoded with DWARF 5, in contrast to DWARF 4 used in Android 9 through 12. Although the actual reduction in binary size can vary depending on different compiler flags, the amount of debug information, and optimization settings, DWARF 5 in general produces smaller and more efficiently parsed binaries especially in large, complex applications. In the next research question we explore the actual symbol growth across the different versions.

*RQ2: Changes in Debugging Symbols:* To better profile the ART library, this research question examines debugging symbol size as an additional metric. We evaluate how architecture and Android version influence the size of debugging symbols. Specifically, given the significant reduction in binary size and the new DWARF 5 encoding between Android 12 and 13/14 across all architectures, how does this impact the size of symbols present in the binary?

The data in Table I shows that symbol count increases almost linearly from Android version 9 to 14 across all architectures, with a slight dip in Android 13. Notably, this growth occurs despite a significant reduction in binary size for Android 13 and 14, highlighting the efficiency of DWARF 5 in encoding rich debug information with a smaller footprint.

Overall, from Android 9 to 14, the symbol count increased by approximately 22% – from 15,698 to 19,115 in ARM64 – indicating the addition of new runtime structures. Android 13 exhibited a modest decrease in symbol count compared to Android 12 (17,743 to 17,275 in ARM64), coinciding with the transition to DWARF 5 encoding. However, Android 14 recorded the highest symbol count across all versions (approximately 19,100 across architectures) while still maintaining a relatively compact binary size.

Additionally, the data reveals consistent DWARF debugging symbol counts across ARM and x86 variants, in both 32-bit and 64-bit implementations for each version. As shown in Table I(b), in Android 14, the symbol count varies by less than 0.1% across all architectures (19,115 for ARM64 vs. 19,097 for x86_32). This consistency suggests that Android's runtime implementation maintains a stable structures across architecture for each version.

This trend characterized by the growth in symbol count and the adoption of more compact encoding underscores the increasing complexity of the Android runtime and suggests that future versions will continue to expand and modularize. As the volume of debug information grows, Android is likely to adopt even more compact encoding strategies and stronger compression techniques to control binary size. While these optimizations enhance runtime performance and storage efficiency, they require forensic tool developers to continuously update their symbol extraction capabilities to support new debug formats, adding to the maintenance burden of symbol-based approaches even when source code, such as AOSP, is available.

> **Findings for Cross-Version Binary Profiling.**
> As Android's runtime continues to grow in complexity, symbol counts are increasing, and more compact encoding and compression strategies will be adopted to manage binary size. While these changes improve performance and efficiency, they require forensic tool developers to continuously design and implement ad-hoc parsers for evolving debug formats, signaling a growing maintenance burden and limitation of symbol-based approaches for memory forensics even in open-source ecosystems like AOSP.

*B. Symbol-Based Structural Evolution Analysis*

This subsection presents the results of the second data analysis method—focused on Symbol-Based Structural Evolution—introduced in Section III. Note that for our architectural analysis, we concentrated on x86_64 architecture because our comparative analysis in Table I(b) demonstrates that structures and their members remain consistent across different architectures for each version, with nearly identical symbol counts. While the actual memory layouts vary due to architecture-specific alignment constraints, the structure layout and relationships remain uniform, making x86_64 an excellent representative of all supported architectures.

*RQ3: Broad ART Structural Evolution:* To examine the evolution of Android's runtime data structures, we analyzed changes in 34 critical structures across consecutive Android versions, from Android 9 to 14. Our analysis reveals 1,067 structural modifications, categorized into three main types: offset changes (shifts in member locations within structures), additions of new members, and removals of existing members. As shown in Table II, offset changes constitute the majority of all modifications, accounting for 89.60% of the total, while member additions and removals constitute 6.37% and 3.66%, respectively. The distribution of these changes across version transitions shows distinct patterns. Android 9 to 10 had the highest concentration of changes with 355 total modifications (312 offset changes, 24 additions, 19 removals), accounting for 33.27% of all changes observed in this study period. This is followed by Android 13 to 14 with 223 modifications (203 offset changes, 11 additions, 9 removals), and Android 12 to 13 with 201 modifications (187 offset changes, 10 additions, 4 removals). The transition between Android 11 and 12 shows the lowest number of modifications with only 109 total changes (98 offset changes, 8 additions, 3 removals), representing a 37.7% decrease from the previous transition period. This pattern suggests a stabilization phase in the runtime architecture following the major restructuring in earlier versions.

Investigating further into the structural differences between Android 9 and 10, we identified significant offset changes across several key data structures, including those related to monitoring, heap management, threads, memory mapping, and allocation structures such as Region and RegionSpace. One noteworthy change was the introduction of new functionality for RegionSpace creation in Android 10. While Android versions before 10 all utilized memmap to create regions without support for Generational Garbage Collection (GGC), Android 10 enhanced the *RegionSpace::create()* function by adding an option to enable Generational Concurrent Copying (use_generational_cc). In traditional GGC, the following assumptions are made: newer objects have shorter lifespans than older objects in memory, newer objects tend to relate to each other and are accessed in close proximity (spatial locality), and it is faster to compact part of the heap than the entire heap[37]. When applied to region space memory allocation, where the heap is divided into equally sized regions and objects are allocated almost sequentially[3], GGC leverages this layout to focus garbage collection on regions containing newer objects (typically referred to as the "young generation"). This approach minimizes the overhead of scanning older regions, where objects are more likely to survive and remain in memory. Furthermore, contiguous allocation within a region enhances performance by leveraging spatial locality and when a region is compacted or collected, related objects can be processed together, reducing the fragmentation and complexity of memory management. Additionally, Android 10 introduced the cyclic region allocation strategy, a significant improvement for memory allocation. According to Androidxref[38], when this option is enabled, new region allocation does not begin at the start of the region space but continues from the last allocated region. This strategy helps detect garbage collection (GC) bugs earlier by reducing immediate region reuse. However, cyclic region allocation also introduces the potential for memory fragmentation at the region level. To mitigate this, the feature is enabled only in debug mode, ensuring it can be tested and validated without adversely affecting production environments. These changes highlight a broader shift in Android from version 10 onward toward enhanced memory management and debugging capabilities. The structural updates supporting these features, particularly in memory allocation and garbage collection, significantly contributed to the observed offset changes in runtime data structures during this transition. For tools that rely on symbol-based analysis, such as **DroidScraper**, these changes in symbol structure composition and layout not only require updates to symbol definitions, but also necessitate substantial modifications to the underlying codebase in order to support Android 10 and later versions.

| Version Transition | Change Type | | | | Total Impact |
|---|---|---|---|---|---|
| | Offset Changes | Member Additions | Member Removals | Structure Removals | |
| Android 9–10 | 312 | 24 | 19 | 1 | 356 |
| Android 10–11 | 156 | 15 | 4 | 0 | 175 |
| Android 11–12 | 98 | 8 | 3 | 0 | 109 |
| Android 12–13 | 187 | 10 | 4 | 3 | 204 |
| Android 13–14 | 203 | 11 | 9 | 0 | 223 |
| **Total** | **956** | **68** | **39** | **4** | **1067** |

TABLE II: Structural Changes Analysis Across Android Versions

*RQ4: In-Depth ART Core Structural Evolution:* The Android Runtime environment relies heavily on three fundamental structures: Runtime, Thread, and Heap. These structures are critical for forensic analysis as they control process execution, thread management, and memory allocation - key aspects needed to reconstruct the system state from the memory image. Our analysis examines both the absolute size changes of these structures and their internal reorganization patterns. The size evolution is particularly significant as it reflects the addition of new capabilities and changes in how the runtime manages system resources. A structure's growth often indicates new functionality or enhanced features, while periods of size stability suggest architectural maturity. Our size evolution analysis reveals distinct growth patterns reflecting the changing roles and maturity of these structures. As shown in Figure 2, the Thread structure demonstrates the most dramatic growth, expanding from 2584 bytes in Android 9 to 6768 bytes in Android 10, a 162% increase. However, this data structure seems to have stabilized since then. On the other hand, the Runtime structure demonstrates the most variable growth over time, expanding and/or shrinking with each new version release. This variability reflects the continuous changes in runtime management capabilities and optimization features across versions. Overall, the Heap structure had the most consistent pattern and size across all the six versions analyzed, signifying the maturity of its design. Our in-depth analysis of the Thread structure across Android versions 9 and 10 reveals that Android 10 introduced a new member, interpreter_cache_, which alone occupies more than 4,000 bytes of memory. This thread-local cache is designed for fast access and supports arbitrary pointer-sized key-value pairs, with the value's interpretation depending on the key[39], thereby providing significant enhancement to the runtime execution efficiency for interpreted code.

The big leap for the Runtime structure was from Android 11 to 12, showing a 62.7% increase. The newly added members include `-boot_class_path_checksums_`, `compat_framework_`, `monitor_timeout_enable_`, `monitor_timeout_ns_`, `is_profileable_`, `madvise_willneed_vdex_filesize_`, `madvise_willneed_odex_filesize_`, `madvise_willneed_art_filesize_`, `deny_art_apex_data_files_`, `force_java_zygote_fork_loop_`, `perfetto_javaheapprof_enabled_`, `metrics_1680`, `metrics_reporter_`, `apex_versions_`, and `app_info_`.

Some of these new members are for compatibility behavior management in the new Android Compatibility Framework. Others are for runtime advice on the expected use of vdex, odex, or art in the near future, security access control for untrustworthy files, and other runtime metrics.

To analyze the distribution and intensity of internal changes, we developed an impact score ranging from 0 to 1.0, where higher scores indicate more significant structural modifications. This score considers three factors: the percentage of members affected by offset changes, the ratio of new or removed members to total members, and the magnitude of size changes relative to the original structure size. For instance, a score of 0.9 indicates that nearly all members were affected by changes, while a score of 0.3 suggests more localized modifications. The heat map visualization in Figure 3 uses these impact scores to illustrate change patterns across versions. For the Runtime structure, the high impact score (0.9) during the Android 9-10 transition indicates major restructuring, with subsequent versions showing decreasing impact scores (0.8 to 0.7) as the structure stabilized. The Thread structure exhibits intense changes (0.8) during the same transition but drops significantly afterward (below 0.4), confirming our size evolution observations. The Heap structure maintains consistently lower impact scores (rarely exceeding 0.5), reflecting its more measured evolution approach.

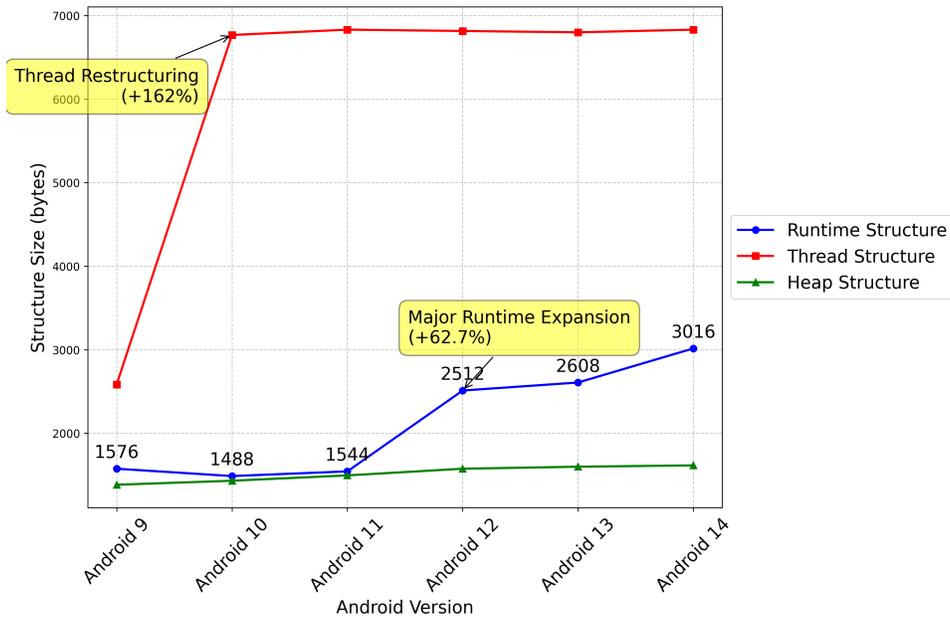

Fig. 2: Evolution of Core Structure Sizes in Android Runtime

The impact score analysis underscores that core runtime structures, such as Runtime and Thread, underwent substantial changes across Android versions. The magnitude and variability of these structural modifications present significant challenges for maintaining symbol-based forensic tools. These findings raise important concerns about the long-term practicality of symbol-based approaches, as they require continuous adaptation to structure layouts making them increasingly difficult to sustain in the context of Android's evolving runtime environment.

*RQ5: Impact of Structural Evolution on Critical Forensic Tasks:* Following our analysis of core structure evolution patterns and their impact scores, we now investigate how these structural changes and their associated member modifications propagate through critical forensic tasks. This analysis is crucial as the architectural changes we observed in Runtime, Thread, and Heap structures, along with the evolution of their internal fields, directly affect the reliability and adaptability of memory analysis. Userland memory forensics, particularly on Android, relies on three fundamental capabilities: thread recovery and enumeration, heap analysis, and object reconstruction. Thread recovery and enumeration serve as the foundation for enabling investigators to track the activity of processes and identify suspicious functionalities by exploring thread states and execution contexts. Heap analysis, on the other hand, aids in the recovery of sensitive runtime data and application artifacts stored in dynamic memory, while object reconstruction enables the retrieval of high-level application data structures and their relationships. These tasks form an interconnected analysis chain where the precision and reliability of the analysis process depend on the accuracy and correctness of the runtime structures and their relationships. Thread enumeration requires parsing the Runtime structure to locate the thread list and traversing Thread structures to understand the process execution state. Heap analysis builds upon thread enumeration by accessing each thread's allocated memory regions through the Heap structure, while object recovery combines information from both threads and heap regions to reconstruct application data. This interdependency has become even more crucial when threads are allocated using TLAB (Thread Local Allocation Buffers). In TLAB, each thread is assigned a specific region in the heap for object allocation. Understanding this thread-allocation feature is crucial in understanding Android object allocation patterns and contexts and how to decode each recovered object.

- **Changes in Thread Recovery and Enumeration:** Thread state analysis has grown increasingly complex across Android versions, primarily due to continuous changes in structure member locations. To enumerate runtime threads for a process we need to find the `thread_­list_` member in the Runtime structure, which provides pointer to the `ThreadList` structure containing the double-linked list of thread objects. This crucial member has shown significant volatility, shifting positions within the Runtime structure across all versions: from offset 512

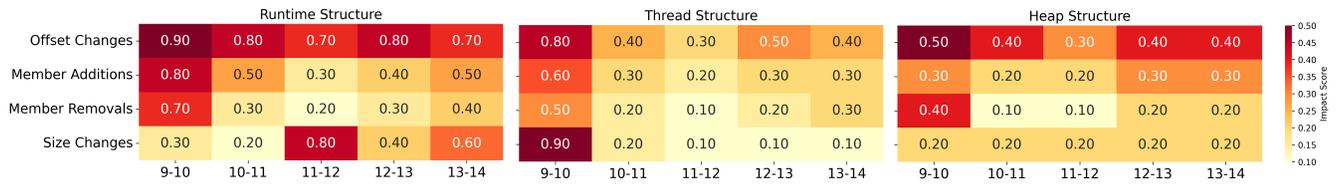
Fig. 3: Structural Changes Impact Analysis Across Components

in Android 9, to 464, 456, 480, 576, and finally 584 in Android 14. This lack of positional stability complicates the fundamental task of locating and traversing thread lists. The complexity of evolution extends beyond the primary thread list pointer. Core thread identification fields such as `name_` and `tid_`, which store essential thread identification information, have undergone location changes in approximately 50% of the versions we analyzed. Similar patterns of positional changes are observed in supporting structures used during thread analysis, including thread state tracking mechanisms. For instance, the `park_state_` member, introduced in Android 10 at offset 60 in `tls_32bit_sized_values` structure, underwent multiple relocations with its new location at offset 44 in Android 14. All these changes reflect the complexity and difficulty in building and maintaining thread recovery and enumeration tools, techniques, and processes that can work across multiple versions.

- **Changes in Heap Analysis:** Heap analysis reveals similar patterns of structural volatility that potentially affect Userland memory forensic capabilities. The `heap_` pointer in the Runtime structure, which provides the critical entry point to heap memory analysis, demonstrates positional instability across versions. Like the thread_list_ pointer, its location shifts unpredictably - moving from offsets 448 → 400 → 392 → 416 → 512 → 512 showing no consistent pattern in its evolution. Within the Heap structure, the `region_space_` member, essential for recovering memory regions and their allocations, has shifted from offset 728 in version 9 to 840 in version 14 accommodating the Android's memory management strategy as discussed in subsection 4.1. These changes are further complicated by the reorganization of the `RegionSpace` structure, where critical members like `num_regions_` that provides information about the number of regions have undergone multiple relocations, impacting the ability to maintain heap analysis over multiple versions.

- **Changes in Object Recovery and Reconstruction:** Object reconstruction represents a particularly critical forensic task in Android Userland memory analysis, given the object-oriented nature of the environment. Application-level forensic artifacts are typically embedded as instance of classes and/or instance or static fields within objects, making accurate object decoding essential for evidence recovery. Prior to Android 13, this task relied heavily on the Object and Class structures, particularly their `ifields_` and other members crucial for decoding object data. While versions 9-12 showed evolution patterns similar to other forensic structures, with member relocations and offset changes, the most significant change occurred in the Android 12-13 transition. As shown in Table II, this transition marked the **removal of fundamental structures** including Object and Class from the `libart.so` library, representing a fundamental shift that requires forensic tools to completely reimagine their approach to object reconstruction.

**Architectural Shift and their forensic implications:** Beyond the structural modifications observed in existing components and their impact on key forensic tasks, we also identified core architectural optimizations—such as the shift from Ahead-of-Time (AOT) compilation to profiler-based Just-In-Time (JIT) compilation—that introduce unique challenges for memory forensics. For instance, during the transition from Android 9 to 10, new structures such as *JitCodeCache* and *ProfilingInfo* were introduced to support the profiler-based JIT compilation. In Android 9 and earlier versions, all *DexFile* instances are embedded within *OATFiles*, which are managed by the *OatFileManager* structure. Forensic tools like Droidscraper recover *DexFile* instances by targeting the *OatFileManager* in the *Runtime* structure, dereferencing the vector it holds to locate each *OATFile*. Once the *OATFile* is identified, the tool then parses it to extract the embedded *DexFile*. However, with the introduction of Android 10, the functionality of the *OatFileManager* structure changed. While it remains part of the runtime, it is only used to hold *BootImage OATFiles* and no longer manages application-specific *OATFiles*. Instead, application Dex files are dynamically compiled and cached. To recover these file instances in Android 10 and later, forensic tools must now trace through a more complex series of structures: starting with the *JitCodeCache* → *ProfilingInfo* → *ArtMethod* → *DexCache* → *DexFile*.

Thus, this transition illustrates why simply updating structure offsets is insufficient for maintaining forensics tools. Even with accurate layout definitions, forensic tools designed for specific version of Android would fail with newer versions without substantial redesign to accommodate the evolving runtime architecture and the new relationships among execution-time structures.

> **Findings for Symbol-Based Structural Evolution Analysis.**

> Our broad analysis of structural modifications in the Android Runtime reveals distinct phases of architectural evolution: significant reorganization driven by memory management enhancements (Android 9–10), stabilization (Android 11–12), and measured evolution (Android 12–14), with offset changes dominating the observed modifications. The core runtime structures exhibit distinct evolution strategies: Runtime shows continuous growth, reflecting expanding capabilities; Thread undergoes a major refactoring phase, followed by structural stability; and Heap follows a consistent, incremental evolution pattern. These structural trends, combined with increasing complexity in forensic tasks such as thread state analysis, heap analysis, and object reconstruction, underscore the challenges in maintaining compatibility across versions. Critically, our findings question the long-term practicality of symbol-based forensic tools, which struggle to keep pace with both fine-grained member relocations and the fundamental architectural shifts observed such as the removal of key structures in Android 13. As structural variability and encoding complexity continue to rise, resilient forensic tools will need to move beyond static symbol reliance and adopt more adaptive analysis methods.

## V. Discussion

The evolution of Android Runtime structures presents challenges that extend far beyond individual structural changes. Our analysis reveals fundamental implications for forensic tool development and maintenance that affect both basic parsing capabilities and complex analysis workflows. The most immediate challenge lies in the stability of structure member locations. Our analysis reveals that 73.2% of surviving structure members (fields that remain present accross version) experienced at least one offset change across versions, creating a moving target for forensic tools. The Runtime structure emerged as particularly volatile, with 89.4% of its members being repositioned at least once. Even thread management structures, which showed relative stability compared to other components, saw 64.7% of their members undergo offset changes. These changes force forensic tools to implement increasingly sophisticated parsing strategies, as a single offset modification can break existing analysis logic.

The introduction of new features such as GGC, cyclic allocation, CompatFramework, and other changes in Android significantly increases the complexity of analyzing userland memory. Additionally, the removal of structures from the ART library `libart.so` presents an even more significant challenge, requiring forensic tools to adapt to distinct scenarios. Our analysis of the Android 9-10 transition demonstrates this through the removal of the DexFile structure, crucial for analyzing DEX (Dalvik Executable) files in memory. This structure enables the investigation of application behavior, method calls, and potential malicious functions by providing access to compiled bytecode and class definitions. Its relocation to the Dex library `libdex.so` required tools to implement additional library parsing capabilities and maintain cross-library references. Similar challenges emerged in the Android 12-13 transition with critical structures like Object and Class, requiring tools to develop new techniques for basic object identification and reconstruction. While our analysis did not observe the removal of individual members critical to forensics, this possibility presents another potential challenge. The removal of fields used for forensic analysis, such as those containing process identifiers, thread states, or security context information, could significantly impact tools' ability to extract crucial evidence, even if the containing structure remains accessible.

These structural changes create cascading effects through different levels of forensic analysis. At the basic level, modifications to structure layouts necessitate updates to parsing logic. More significantly, these changes affect higher-level analysis capabilities, exemplified by the impact on string constant recovery and method implementation analysis following DexCache structure modifications. The complexity extends to interrelated analysis tasks. Object reconstruction requires multiple fallbacks and inference mechanisms to handle cases where traditional parsing methods fail. The introduction of new security features and runtime optimizations adds another layer of complexity, requiring tools to adapt to both structural changes and evolving runtime behaviors.

Thus, putting all the findings together, this measurement study suggest that symbol-based approaches, while historically foundational, are increasingly unsustainable for memory forensic analysis. The rate of structural evolution through offset shifts, structure relocations, and the introduction of architectural indirection means that simply updating symbols or offsets is no longer sufficient. Android's runtime is constantly been reengineered for resilience, performance, and modularity. Future tools must adopt version-aware, flexible analysis methodologies that preserve interpretability, reliability, and evidentiary soundness. Although our study focuses on Android, these challenges are symptomatic of broader shifts across modern software platforms. Thus, our work serves not only as a structural analysis of Android's evolution but also as a broader call to reimagine memory forensic tooling in an era of continuous architectural change.

## VI. Recommendations and Future Directions

To address the growing challenges posed by structural evolution in Android, we recommend that memory forensic tools move beyond sole reliance on static symbol extraction. However, we do not advocate abandoning symbol-based methods. Instead, we propose hybrid approaches that preserve the validation, traceability, and reliability inherent in symbol-based techniques while improving adaptability across versions.

One promising direction is to integrate symbolic execution as a complementary analysis strategy. Symbolic execution can infer how data structures are traversed and accessed in practice, capturing semantic relationships between fields that purely pattern-based methods miss. The work of Maggio et al. demonstrates this potential in a proof-of-concept known as

Seance, which applies symbolic execution to identify structural differences in symbol definitions across versions of binaries or libraries [40]. Rather than relying solely on manual diffing or fragile offset comparisons, Seance analyzes how symbols are used in compiled code to infer where structures have changed, offering a version-aware and behaviorally informed mechanism for updating structure definitions. However, Seance has important limitations that highlight the need for further development. As presented at Black Hat 2022 [41], Seance failed to detect changes in references to global variables, reporting functions as unchanged despite significant shifts in global state. This discrepancy broke tools that relied on those global references, illustrating the fragility of current symbolic execution approaches without robust symbol tracking or validation. To address these limitations, our future work will develop enhanced techniques that combine directed symbolic execution with memory access pattern analysis. By guiding symbolic execution along specific code paths that lead to data structure traversal, we can systematically infer structure members and their relative offsets through analysis of memory access patterns during execution. This approach will provide a more robust foundation for automatically inferring structure layouts while maintaining the validation strength necessary for forensic applications.

Pattern-matching techniques, such as those in [4, 5, 29], offer an additional alternative for identifying in-memory structures. However, without symbolic or semantic insight, they may yield incomplete or fragmented structures – especially in Android, where data layouts are complex and interdependent. Symbolic execution fills this gap by offering both interpretability and validation, two critical requirements for forensic soundness.

To operationalize these approaches, we recommend that tools adopt version-aware parsing architectures, where structure definitions are modular and parsing logic is decoupled from fixed layouts. Tools should maintain version-specific maps while supporting shared parsing logic, easing updates and reducing maintenance overhead. Additionally, critical analysis paths—such as locating Runtime or Thread structures—should be built with redundancy, offering multiple discovery methods, including fallback to memory scanning or behavioral pattern tracing.

## VII. Conclusion

This study presents the first comprehensive empirical analysis of structural evolution in the Android Runtime (ART) and its implications for userland memory forensics. Through a cross-version examination of critical runtime structures across Android versions 9 to 14 for four different architectures, we demonstrate that structural changes particularly offset shifts, structure relocations, and removals are not only widespread but also significantly disruptive to traditional forensic workflows. Our findings reveal that 73.2% of persistent structure members changed positions at least once, and that core components such as Runtime, Thread, and Heap exhibit distinct and evolving structural patterns that challenge the assumptions of static symbol-based analysis. Beyond individual structural changes, architectural transformations such as the adoption of generational garbage collection, JIT compilation, and modular runtime components further complicate structure interpretation and symbol stability. These changes underscore the growing fragility of memory forensic tools that rely solely on static symbol definitions.

We argue that while symbol-based analysis remains invaluable for its reliability and validation, it is no longer sufficient on its own and in fact these tools risk obsolescence without continual adaptation. Hence, future forensic tooling must incorporate version-aware, flexible, and semantically informed methodologies that can adapt to evolving structure layouts without sacrificing evidentiary soundness. Hybrid approaches that combine symbolic reasoning and selective ground-truth validation offer a promising path forward, though significant research and development are needed to operationalize these methods for practical forensic use. Ultimately, as software ecosystems grow more dynamic and modular, the sustainability of memory forensic methods will depend on their ability to match the systems they aim to analyze—adaptable, resilient, and rigorously validated.

## VIII. Acknowledgments

This work was supported by the Department of Homeland Security (DHS) under the Criminal Investigations and Newtork Analysis Center (CINA) Grant No. E206378P

## References

[1] V. Foundation, "Volatility framework: Volatile memory artifact extraction utility framework." https://github.com/volatilityfoundation/volatility, 2017. Accessed: Jan. 11, 2025.

[2] M. Cohen, "Rekall memory forensic framework." https://github.com/google/rekall, 2014. Accessed: Apr. 24, 2025.

[3] A. Ali-Gombe, S. Sudhakaran, A. Case, and G. G. Richard III, "{DroidScraper}: A tool for android {In-Memory} object recovery and reconstruction," in *22nd International Symposium on Research in Attacks, Intrusions and Defenses (RAID 2019)*, pp. 547–559, 2019.

[4] A. Oliveri, M. Dell'Amico, and D. Balzarotti, "An os-agnostic approach to memory forensics," in *NDSS 2023, Network and Distributed System Security Symposium, 27 February-3 March 2023, San Diego, CA, USA*, Internet Society, 2023.

[5] F. Franzen, T. Holl, M. Andreas, J. Kirsch, and J. Grossklags, "Katana: Robust, automated, binary-only forensic analysis of linux memory snapshots," in *Proceedings of the 25th International Symposium on Research in Attacks, Intrusions and Defenses*, pp. 214–231, 2022.

[6] "Anonymous github repository for android runtime structures data (versions 9–14) and source code of study framework to support memory forensics research." https://anonymous.4open.science/r/ART_Evoltution_

MForensic-220D/README.md, 2025. Accessed: Apr. 24, 2025.

[7] A. Case, "Memory analysis of the dalvik (android) virtual machine," 2011.

[8] H. Macht, "Live memory forensics on android with volatility," *Friedrich-Alexander University Erlangen-Nuremberg*, 2013.

[9] B. Saltaformaggio, R. Bhatia, Z. Gu, X. Zhang, and D. Xu, "Guitar: Piecing together android app guis from memory images," in *Proceedings of the 22nd ACM SIGSAC Conference on Computer and Communications Security*, pp. 120–132, 2015.

[10] B. Saltaformaggio, R. Bhatia, Z. Gu, X. Zhang, and D. Xu, "Vcr: App-agnostic recovery of photographic evidence from android device memory images," in *Proceedings of the 22nd ACM SIGSAC Conference on Computer and Communications Security*, pp. 146–157, 2015.

[11] A. M. M. Soares and R. T. de Sousa Jr, "A technique for extraction and analysis of application heap objects within android runtime (art).," in *ICISSP*, pp. 147–156, 2017.

[12] R. Bhatia, B. Saltaformaggio, S. J. Yang, A. I. Ali-Gombe, X. Zhang, D. Xu, and G. G. Richard III, "Tipped off by your memory allocator: Device-wide user activity sequencing from android memory images.," in *NDSS*, 2018.

[13] Android Open Source Project (AOSP), "Gc debugging in android runtime." https://source.android.com/docs/core/runtime/gc-debug, 2025. Accessed: Jan. 29, 2025.

[14] Android Open Source Project (AOSP), "Improvements in android runtime." https://source.android.com/docs/core/runtime/improvements, 2025. Accessed: Jan. 29, 2025.

[15] A. Ali-Gombe, A. Tambaoan, A. Gurfolino, and G. G. Richard III, "App-agnostic post-execution semantic analysis of android in-memory forensics artifacts," in *Proceedings of the 36th Annual Computer Security Applications Conference*, pp. 28–41, 2020.

[16] S. Sudhakaran, A. Ali-Gombe, A. Orgah, A. Case, and G. G. Richard, "Ampledroid recovering large object files from android application memory," in *2020 IEEE International Workshop on Information Forensics and Security (WIFS)*, pp. 1–6, IEEE, 2020.

[17] A. Case and G. G. Richard III, "Memory forensics: The path forward," *Digital investigation*, vol. 20, pp. 23–33, 2017.

[18] N. L. Petroni Jr, A. Walters, T. Fraser, and W. A. Arbaugh, "Fatkit: A framework for the extraction and analysis of digital forensic data from volatile system memory," *Digital Investigation*, vol. 3, no. 4, pp. 197–210, 2006.

[19] J. Sylve, A. Case, L. Marziale, and G. G. Richard, "Acquisition and analysis of volatile memory from android devices," *Digital Investigation*, vol. 8, no. 3-4, pp. 175–184, 2012.

[20] N. Lewis, A. Case, A. Ali-Gombe, and G. G. Richard III, "Memory forensics and the windows subsystem for linux," *Digital Investigation*, vol. 26, pp. S3–S11, 2018.

[21] M. H. Ligh, A. Case, J. Levy, and A. Walters, *The art of memory forensics: detecting malware and threats in windows, linux, and Mac memory*. John Wiley & Sons, 2014.

[22] A. Case, M. M. Jalalzai, M. Firoz-Ul-Amin, R. D. Maggio, A. Ali-Gombe, M. Sun, and G. G. Richard III, "Hooktracer: A system for automated and accessible api hooks analysis," *Digital Investigation*, vol. 29, pp. S104–S112, 2019.

[23] F. Block and A. Dewald, "Linux memory forensics: Dissecting the user space process heap," *Digital Investigation*, vol. 22, pp. S66–S75, 2017.

[24] F. Block and A. Dewald, "Windows memory forensics: Detecting (un) intentionally hidden injected code by examining page table entries," *Digital Investigation*, vol. 29, pp. S3–S12, 2019.

[25] D. Oygenblik, C. Yagemann, J. Zhang, A. Mastali, J. Park, and B. Saltaformaggio, "{AI} psychiatry: Forensic investigation of deep learning networks in memory images," in *33rd USENIX Security Symposium (USENIX Security 24)*, pp. 1687–1704, 2024.

[26] Z. Lin, J. Rhee, X. Zhang, D. Xu, and X. Jiang, "Siggraph: Brute force scanning of kernel data structure instances using graph-based signatures.," in *Ndss*, 2011.

[27] Q. Feng, A. Prakash, M. Wang, C. Carmony, and H. Yin, "Origen: Automatic extraction of offset-revealing instructions for cross-version memory analysis," in *Proceedings of the 11th ACM on Asia conference on computer and communications security*, pp. 11–22, 2016.

[28] M. I. Cohen, "Characterization of the windows kernel version variability for accurate memory analysis," *Digital Investigation*, vol. 12, pp. S38–S49, 2015.

[29] Z. Qi, Y. Qu, and H. Yin, "Logicmem: Automatic profile generation for binary-only memory forensics via logic inference.," in *NDSS*, 2022.

[30] F. Pagani and D. Balzarotti, "Autoprofile: Towards automated profile generation for memory analysis," *ACM Transactions on Privacy and Security*, vol. 25, no. 1, pp. 1–26, 2021.

[31] B. Saltaformaggio, Z. Gu, X. Zhang, and D. Xu, "{DSCRETE}: Automatic rendering of forensic information from memory images via application logic reuse," in *23rd USENIX Security Symposium (USENIX Security 14)*, pp. 255–269, 2014.

[32] W. Song, H. Yin, C. Liu, and D. Song, "Deepmem: Learning graph neural network models for fast and robust memory forensic analysis," in *Proceedings of the 2018 ACM SIGSAC Conference on Computer and Communications Security*, pp. 606–618, 2018.

[33] S. Zhang, X. Meng, and L. Wang, "An adaptive approach for linux memory analysis based on kernel code reconstruction," *EURASIP Journal on Information Security*, vol. 2016, no. 1, p. 14, 2016.

[34] Android Open Source Project (AOSP), "Requirements for setting up the android source." https://source.android.com/docs/setup/start/requirements, 2025. Accessed: Jan.

29, 2025.

[35] Volatility Foundation, "dwarf2json: A tool for converting dwarf debugging information to json." https://github.com/volatilityfoundation/dwarf2json, 2025. Accessed: Jan. 28, 2025.

[36] E. Bendersky, "pyelftools: A pure-python library for parsing elf and dwarf files." https://github.com/eliben/pyelftools, 2025. Accessed: Jan. 29, 2025.

[37] M. Documentation, "Fundamentals of garbage collection." https://learn.microsoft.com/en-us/dotnet/standard/garbage-collection/fundamentals, 2025. Accessed: Jan. 11, 2025.

[38] Android Open Source Project (AOSP), "region_space.h - android 10 reference." https://xrefandroid.com/android-10.0.0_r47/xref/art/runtime/gc/space/region_space.h, 2025. Accessed: Jan. 11, 2025.

[39] Android Open Source Project (AOSP), "interpreter_cache.h - android 10 reference." https://xrefandroid.com/android-10.0.0_r47/xref/art/runtime/interpreter/interpreter_cache.h, 2025. Accessed: Jan. 11, 2025.

[40] R. D. Maggio, A. Case, A. Ali-Gombe, and G. G. Richard, "Seance: Divination of tool-breaking changes in forensically important binaries," *Forensic Science International: Digital Investigation*, vol. 37, p. 301189, 2021.

[41] A. Case, "New memory forensics techniques to defeat device monitoring malware." [Online]. Available: https://i.blackhat.com/USA-22/Wednesday/US-22-Case-New-Memory-Forensics-Techniques-to-Defeat-Device-Monitoring-Malware-wp.pdf, 2022. [Accessed: Jan. 29, 2025].